\documentclass{aa}
\usepackage{graphicx}
\usepackage{float}

\begin{document} 

\title{\textsc{ShapePipe}: A modular weak-lensing processing and analysis pipeline}

\author{S. Farrens\inst{\ref{cosmostat}} \thanks{email: samuel.farrens@cea.fr}
    \and A. Guinot\inst{\ref{apc}}
    \and M. Kilbinger\inst{\ref{cosmostat}}
    \and T. Liaudat\inst{\ref{cosmostat}}
    \and L. Baumont\inst{\ref{cosmostat}} 
    \and X. Jimenez\inst{\ref{ens}}
    \and A. Peel\inst{\ref{epfl}}
    \and A. Pujol\inst{\ref{cosmostat}}
    \and M. Schmitz\inst{\ref{nice}}
    \and J.-L. Starck\inst{\ref{cosmostat}}
    \and A. Z. Vitorelli\inst{\ref{cosmostat}}
}

\institute{AIM, CEA, CNRS, Universit\'{e} Paris-Saclay, Universit\'{e} Paris Diderot,           Sorbonne Paris Cit\'{e}, F-91191 Gif-sur-Yvette, France\label{cosmostat} 
    \and Universit\'{e} de Paris, CNRS, Astroparticule et Cosmologie, F-75013 Paris, France\label{apc}
    \and Universit\'{e} Paris-Saclay, CNRS, ENS Paris-Saclay, Centre Borelli, 91190, Gif-sur-Yvette, France \label{ens}
    \and Institute of Physics, Laboratory of Astrophysics, Ecole Polytechnique F\'{e}d\'{e}rale de Lausanne (EPFL), Observatoire de Sauverny, 1290 Versoix, Switzerland\label{epfl}
    \and Universit\'{e} C\^{o}te d’Azur, Observatoire de la C\^{o}te d’Azur, CNRS, Laboratoire Lagrange, Bd de l’Observatoire, CS 34229, 06304 Nice Cedex 4, France\label{nice}
} 

\date{}
 
\abstract{
We present the first public release of \textsc{ShapePipe}, an open-source and modular weak-lensing measurement, analysis, and validation pipeline written in Python. We describe the design of the software and justify the choices made. We provide a brief description of all the modules currently available and summarise how the pipeline has been applied to real Ultraviolet Near-Infrared Optical Northern Survey data. Finally, we mention plans for future applications and development. The code and accompanying documentation are publicly available on GitHub.
}

\keywords{Gravitational lensing: weak -- Methods: data analysis}

\titlerunning{ShapePipe}
\authorrunning{S. Farrens et al.}

\maketitle

%-----------------------------------------------------------------------
% INTRODUCTION
%-----------------------------------------------------------------------

\section{Introduction}

Weak gravitational lensing, the apparent distortion of the shapes of galaxies caused by the bending of light by mass along the line of sight, has been demonstrated to be a powerful probe of cosmology \citep{kilbinger:15, mandelbaum:18b}. However, numerous steps are required in order to go from raw survey data to competitive constraints on cosmological parameters. Recent surveys, such as the Canada-France-Hawai’i Telescope Legacy Survey \citep{erben:13}, the Hyper Suprime Cam (HSC) survey \citep{mandelbaum:18a}, the Kilo-Degree Survey \citep{Kuijken:19}, and the Dark Energy Survey \citep[DES;][]{gatti:21}, have carried out detailed weak-lensing analyses. Upcoming surveys, such as {\it Euclid} \citep{laureijs:11} and the Vera C. Rubin Observatory Legacy Survey of Space and Time \citep[LSST;][]{ivezic:19}, will also aim to tighten cosmological constraints with weak lensing. It is clear that weak lensing remains at the forefront of cosmological studies, and hence tools for weak-lensing analysis will remain in demand for years to come.

There are various pipelines available for simulating weak-lensing data \citep[e.g.][]{kiessling:11, harnois:15, petri:16}. Various data processing pipelines also exist \citep[e.g.][]{bosch:18}, but these are generally very survey-specific and often private. We believe this leaves an ample gap in the market for an open-source end-to-end weak-lensing pipeline.

We present the first public release of \textsc{ShapePipe}, an open-source and modular weak-lensing measurement, analysis, and validation pipeline written in Python. An earlier version of this pipeline was applied to real survey data in \citet{guinot:22} with promising results. 

The current version of \textsc{ShapePipe} starts with reduced survey images and ends by providing shear measurements along with all of the information required for calibration. It includes a novel point spread function (PSF) modelling technique and various PSF validation tools. The code has been designed to facilitate the inclusion of new or improved processing steps to adapt to advances made in the coming years.

This paper is structured as follows. In the following section we describe the design of the software along with efforts we have made to ensure reproducibility and to encourage external contributions. In Sect.~\ref{sec:wl_pipeline} we describe the various processing elements that comprise the current version of the pipeline. In Sect.~\ref{sec:applications} we summarise how the pipeline has been implemented thus far and what we have in mind with regards to future applications. Section~\ref{sec:future} mentions our plans for future releases, and we end with some concluding remarks.

%-----------------------------------------------------------------------
% SOFTWARE DESIGN
%-----------------------------------------------------------------------

\section{Software design}

\subsection{Modular architecture}

A key feature dictating the design of \textsc{ShapePipe} was flexibility. Weak gravitational lensing is a rapidly evolving field, with new analysis techniques and tools emerging regularly. A hard-coded weak-lensing pipeline would therefore have extremely limited applications and would become quickly outdated. It was therefore essential for us that a new pipeline be able to adapt in a way that was both maintainable and allowed the easy inclusion of new tools.

The \textsc{ShapePipe} package comprises two principle sub-packages: \texttt{pipeline} and \texttt{modules}. The \texttt{pipeline} sub-package manages job handling, logging, configuration, argument parsing, dependency handling, and the reading and writing of Flexible Image Transport System (FITS) files. This can effectively be seen as the core of \textsc{ShapePipe}. The \texttt{modules} sub-package consists of a series of what we refer to as `pipeline modules'\footnote{Not to be confused with Python modules.}, which handle all of the data processing and analysis steps that make up a weak-lensing pipeline (see Sect.~\ref{sec:wl_pipeline}).

Pipeline modules are simply defined by writing a `module runner', which is a standardised \textsc{ShapePipe} script that defines what a given module should do. This could be as simple as calling an executable binary with some options or writing a full implementation in Python. In either case, once a module runner is in place, \textsc{ShapePipe} can call this module in conjunction with any other. Inputs can be automatically passed or taken from modules, making it possible to chain a series of operations with relative ease.

This architecture will allow us to quickly add new modules to \textsc{ShapePipe} as new tools come onto the market without disrupting the existing pipeline structure. It also means we can more easily rearrange or adapt the existing tools to new datasets.

\textsc{ShapePipe} pipelines are launched via a command line execution script. Significant effort has been put into making this script user-friendly and flexible and providing useful and verbose logging information in order to track the success or failure of any given process.

\subsection{Parallel computation}

\textsc{ShapePipe} was developed with the assumption that most weak-lensing survey data can be divided into tiles or patches, each of which can be processed independently. Therefore, \textsc{ShapePipe} distributes jobs in an embarrassingly parallel way using \textsc{Joblib} \citep{joblib:20} for multi-core machines, and \textsc{MPI for Python} \citep{dalcin:05, dalcin:08, dalcin:11} on large computer clusters. This maximises the amount of data that can be processed simultaneously with the number of computer cores available to the user without having to parallelise each of the processing steps independently. Modules can also be configured to run in serial mode if the associated task requires access to all of the data, for example, for recombining tiles or patches.

\subsection{Reproducibility}

The ability to reproduce scientific results is a growing and continuing concern within academia. This is a particularly challenging problem in the context of a weak-lensing pipeline as there are numerous data treatment and processing steps, often involving various third-party tools that evolve over time. We have attempted to mitigate these problems as best as possible.

 Firstly, we provide a dedicated \textsc{ShapePipe} development and processing environment. This consists of a \textsc{Conda}\footnote{\url{https://docs.conda.io/}} environment where all Python packages, compilation tools (such as \textsc{CMake}), and other third-party software packages (see Appendix~\ref{sec:software}) have fixed versions for a given \textsc{ShapePipe} release. This also extends to all of the modules that make up the weak-lensing pipeline (see Appendix~\ref{sec:modules}). This not only ensures consistency across different platforms on which \textsc{ShapePipe} is installed, but also isolates the tools used for processing from those already installed on a given platform (e.g. a different version of Python).
 
 Secondly, we have built in an extensive logging system that tracks the versions of all the packages and modules used for a given run of \textsc{ShapePipe}, along with all the configuration options used. The logging system catches and reports all errors raised by each module run. 
 
 These measures should allow users to reprocess a given dataset in virtually identical conditions. However, we will continue to make reproducibility a priority for future releases of \textsc{ShapePipe}.

\subsection{Open-source development}

\textsc{ShapePipe} is a fully open-source project. All of the code is publicly available on GitHub\footnote{\url{https://github.com/CosmoStat/shapepipe}} along with extensive documentation\footnote{\url{https://CosmoStat.github.io/shapepipe/}}. We provide example configuration files demonstrating how to use the various tools in the context of our application use cases (see Sect.~\ref{sec:applications}), and we additionally provide an extensive step-by-step tutorial for running \textsc{ShapePipe} for processing Ultraviolet Near-Infrared Optical Northern Survey (UNIONS) data (see Sect.~\ref{sec:unions}).

The GitHub repository is configured with template issues and pull requests, and includes detailed contribution guidelines. These features are intended to encourage and facilitate contributions from the wider community. Finally, we have also included a comprehensive code of conduct to ensure that contributors can interact in a safe and supportive environment.

%-----------------------------------------------------------------------
% WEAK-LENSING PIPELINE
%-----------------------------------------------------------------------

\section{Weak-lensing pipeline}
\label{sec:wl_pipeline}

In this section we briefly describe the primary pipeline modules that are included in the first public release of \textsc{ShapePipe} in the context of the weak-lensing processing steps they implement. A full list of the modules currently available is provided in Appendix~\ref{sec:modules}.

\subsection{Masking}

The current masking procedure in \textsc{ShapePipe} takes inspiration from the \textsc{THELI} pipeline \citep{erben:05, schirmer:13} to mask out bright stars, diffraction spikes, Messier objects, and border regions. The \texttt{mask} module uses the \textsc{cdsclient} software to download a catalogue of bright reference stars from version 2.2 of the Guide Star Catalog \citep[GSC;][]{gsc:01}. Alternatively, a star catalogue available on disk (with the same format as the GSC) can also be used, which is particularly relevant for cluster compute nodes that do not allow internet access.

\begin{figure}[ht]
    \centering
    \includegraphics[width=0.49\textwidth]{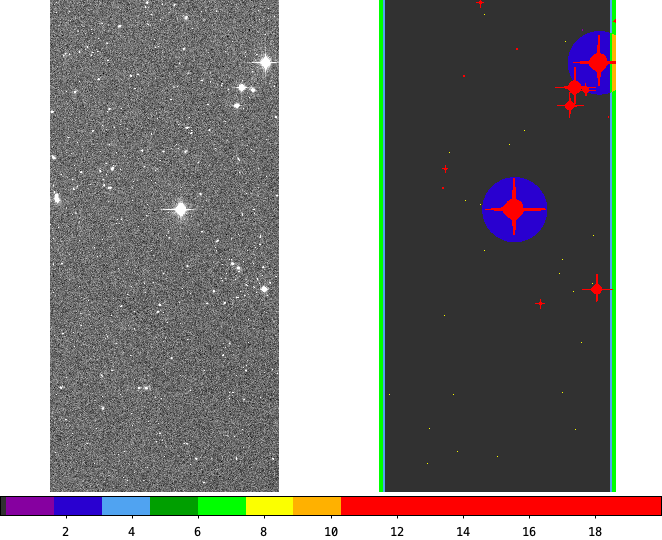}
    \caption{Example of a mask produced by the ShapePipe \texttt{mask} module on a patch of UNIONS data. The colour bar shows the integer mask pixel values.}
    \label{fig:mask_example}
\end{figure}

The mask itself is generated using \textsc{WeightWatcher} \citep{marmo:08}. An example of a mask produced using this module on a patch of UNIONS data is shown in Fig.~\ref{fig:mask_example}.

\subsection{Source detection}

The identification of stars and galaxies in the survey images is performed using the community standard \textsc{SExtractor} \citep{bertin:96}. \textsc{ShapePipe} allows modules such as \textsc{SExtractor} to be run multiple times with different configuration options, meaning that stars and galaxies can be targeted in different ways depending on the need. For example, one run can be made to target a large sample of galaxies for shape measurement, while another run can target a sample of stars for modelling the PSF.

\subsection{Source selection}

\begin{figure*}[ht]
    \centering
    \includegraphics[width=0.49\textwidth]{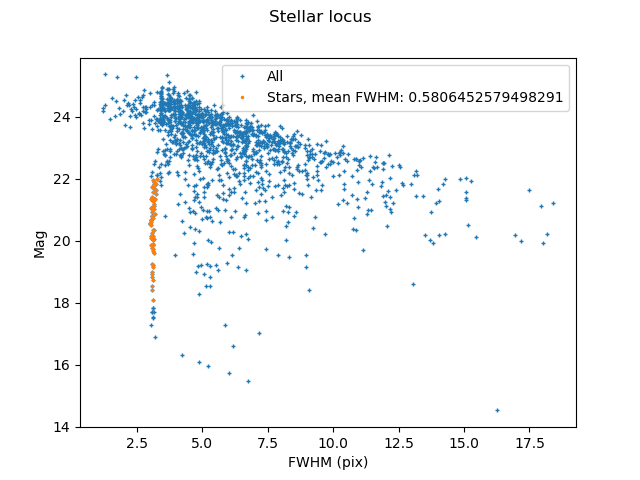}
    \includegraphics[width=0.49\textwidth]{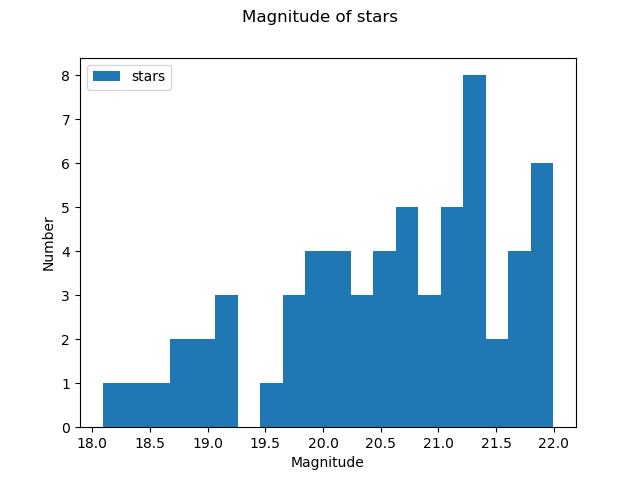}
    \includegraphics[width=0.49\textwidth]{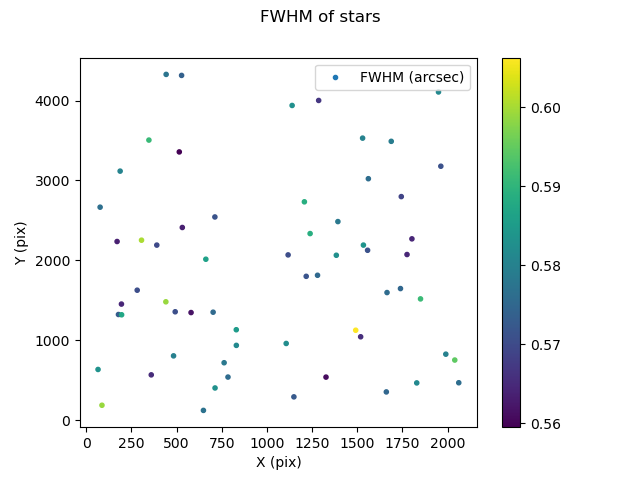}
    \includegraphics[width=0.49\textwidth]{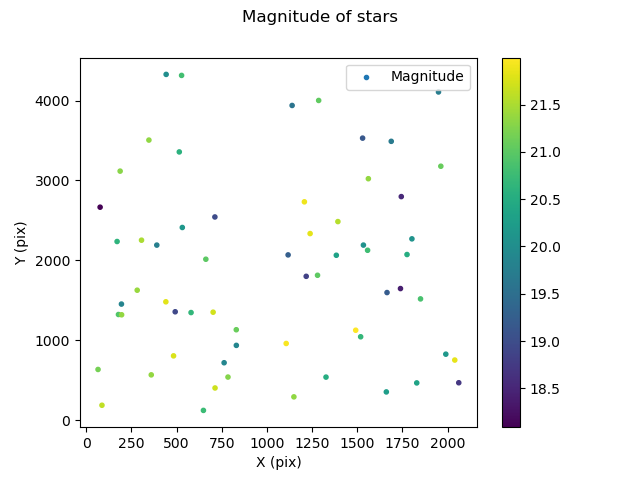}
    \caption{Example of a refined star selection obtained with the \texttt{setools} module on a patch of UNIONS data. All of the plots shown are generated automatically by \textsc{ShapePipe}. \emph{Top left}: Stellar locus (orange points) in the size-magnitude plane.  \emph{Top right}: Histogram of the magnitude of the objects. \emph{Bottom left}: Size of objects in the field. \emph{Bottom right}: Magnitude of objects in the field.}
    \label{fig:star_selection}
\end{figure*}

The star and galaxy samples used for shape measurement can be refined using several \textsc{ShapePipe} modules. One module implements the spread model method proposed by \citet{desai:12} and \citet{mohr:12}, which tries to determine if a given object is more point-like or rather more extended in comparison with the PSF. Another module, called \texttt{setools}, can apply arbitrary cuts to the objects detected by \textsc{SExtractor}. Figure~\ref{fig:star_selection} shows some example plots of a refined star sample obtained after running the \texttt{setools} module in \textsc{ShapePipe}.

\subsection{PSF model}
\label{sec:psf}

\textsc{ShapePipe} includes two independent methods for modelling the PSF of the survey. The first method is the community standard \textsc{PSFEx} \citep{bertin:11}, which was developed to be used in conjunction with \textsc{SExtractor}. \textsc{PSFEx} produces independent models with polynomial variations of the PSF for each charge-coupled device (CCD) in the field and each exposure. \textsc{ShapePipe} includes an additional module for interpolating the \textsc{PSFEx} PSF model from the positions of the star sample to the positions of the galaxy sample.
 
The second method is the multi-CCD (\textsc{MCCD}) approach of \citet{liaudat:21}. \textsc{MCCD} builds a model of the PSF for the entire focal plane, capturing both variations that span several CCDs and those that are localised on individual CCDs. As with \textsc{PSFEx}, there is a  \textsc{ShapePipe} module for interpolating MCCD PSF models to galaxy positions. \textsc{ShapePipe} allows either or both methods to be run and subsequently used for the galaxy shape measurement.

\subsection{PSF validation tools}

\textsc{ShapePipe} includes a variety of tools, such as plots of the PSF residuals and $\rho$-statistics \citep{rowe:10,jarvis:16}, for validating the quality of the PSF model. See \citet{guinot:22} for an example of how these PSF validation tools have be used for the analysis of UNIONS data.

\subsection{Shape measurement}

Galaxy shape measurement is currently performed using \textsc{NGMIX} \citep{sheldon:15}, which has been extensively applied and tested on DES data \citep{jarvis:16,zuntz:18,gatti:21}. At present, \textsc{NGMIX} is implemented in \textsc{ShapePipe} such that galaxy shapes are obtained from a fit of a Gaussian profile from Galsim \citep{rowe:15}. The \textsc{NGMIX} module can be run with either of the PSF modelling methods described in Sect.~\ref{sec:psf}. Additional \textsc{NGMIX} features and alternative shape measurement techniques are likely to be made available in future releases of \textsc{ShapePipe}.

\subsection{Calibration}

The primary shear calibration method currently available in \textsc{ShapePipe} is the so-called metacalibration framework \citep{huff:17, sheldon:17}. Metacalibration is also implemented via the \textsc{NGMIX} module. This module does not directly provide calibrated shear measurements, but does provide all of the calibration images needed to calculate the shear response.

\subsection{Miscellaneous}

In addition to the essential and generic weak-lensing pipeline processing tools described in the preceding subsections, \textsc{ShapePipe} includes various bookkeeping and file management tools. Some of these tools have been designed specifically for handling UNIONS data, but could be extremely useful for anyone wishing to reprocess these data. For example, \textsc{ShapePipe} includes a module for matching objects detected by the pipeline with external catalogues and the possibility to generate random catalogues.

%-----------------------------------------------------------------------
% APPLICATIONS
%-----------------------------------------------------------------------

\section{Applications}
\label{sec:applications}

\subsection{UNIONS}
\label{sec:unions}

The primary focus and initial motivation for the development of \textsc{ShapePipe} has been the analysis of UNIONS data. UNIONS is a collaboration between the Canada-France Imaging Survey (CFIS), Pan-STARRS, and the Wide Imaging with Subaru HSC of the Euclid Sky. One of the main objectives of UNIONS is to provide ground-based multi-band imaging data ($u,g,r,i$, and $z$ bands collectively; see \citealt{ibata:17a} for details on the $u$ band) of the northern sky for {\it Euclid}. High quality ground-based photometry is fundamental for accurate photometric redshift estimation and is therefore critical to the success of the {\it Euclid} mission. In addition to this, there are various ongoing scientific studies with UNIONS data on stars and Milky Way physics \citep{ibata:17b, fantin:19, fantin:21, thomas:18, thomas:19b, thanjavur:21}, globular clusters \citep{thomas:20, jensen:21}, dwarf galaxies \citep{thomas:19a}, galaxy evolution and mergers \citep{bickley:21}, clusters of galaxies \citep{roberts:22, spitzer:21}, high redshift galaxies and active galactic nuclei \citep{ellison:19}, PSF modelling \citep{liaudat:21}, and gravitational lensing \citep{guinot:22, chan:21, savary:21}.

\citet{guinot:22} applied a preliminary version of \textsc{ShapePipe} to 1~700 deg$^2$ of CFIS $r$-band data, obtaining shapes for around 40 million galaxies. The $r$ band of CFIS is particularly well suited to weak-lensing studies given the average seeing of 0.65 and that it is complete down to magnitude 24.5. \citet{guinot:22} demonstrate that \textsc{ShapePipe} is capable of processing a large fraction of the sky (on multiple computing platforms) and achieving promising results. Each of the $\sim 6800$ images of size $0.5 \times 0.5 {\rm deg}^2$ was processed independently. The analysis of an image was performed on a virtual machine (VM) on the Canadian Advanced Network for Astronomical Research\footnote{\url{https://www.canfar.net/}}. Where possible, individual processing steps were run in parallel on eight cores of the VM. The processing time of one image was around $4$ hours. On average, $200$ VMs were used in parallel, resulting in a total wall-clock run time of two weeks. The final shape catalogue was around $120$ Gb in size. Despite the limitations of working with a single photometric band, \citet{guinot:22} found: no B-mode signal, a good correlation between E-mode peaks and Planck clusters, and an overall low level of systematic bias.

Work is currently ongoing to process a larger  CFIS area (3~500 deg$^2$) with a more recent version of \textsc{ShapePipe} (Kilbinger et al., in prep). This work takes advantage of some of the state-of-the-art features available in \textsc{ShapePipe}, such as full focal plane modelling of the PSF using \textsc{MCCD}. The resulting catalogue will contain around 100 million galaxy shapes and will constitute one of the largest weak-lensing catalogues created to date. In addition to this, several follow-up scientific studies are planned (some are already underway) using \textsc{ShapePipe}-processed UNIONS data, such as 3x2 point analysis, peak statistics \citep{aycoberry:22}, and galaxy cluster lensing.

\subsection{Other surveys}
\label{sec:surveys}

No concrete plans have been made at this stage for applying \textsc{ShapePipe} to other datasets. Despite designing the architecture to be as flexible as possible, significant work would still need to be carried out to appropriately tune all of the processing steps to a different survey. Large cosmological surveys such as the DES, {\it Euclid}, and the Vera C. Rubin Observatory LSST already benefit from dedicated weak-lensing analysis pipelines highly tuned to the specific needs of each survey. However, independent reprocessing of some fraction of these datasets could be valuable for cross-validation and comparison. Additionally, some smaller surveys lack the workforce necessary to build a new weak-lensing pipeline, or indeed to adapt an existing pipeline to their data. Either or both of these use cases might be well suited for future \textsc{ShapePipe} applications.

%-----------------------------------------------------------------------
% Future Releases
%-----------------------------------------------------------------------

\section{Future releases}
\label{sec:future}

We plan to provide regular patch and minor releases of \textsc{ShapePipe} to fix bugs and make general improvements to ensure good maintenance of the code. We additionally plan one major release on a roughly yearly basis in which dependency versions will be updated and new or improved processing steps included. Specific improvements and additions that we currently have in mind include improved masking (new tools and use of Gaia data), metadetection, the handling of blended galaxies \citep[see the preliminary work of][]{farrens:22}, and the full handling of multi-band data.

%-----------------------------------------------------------------------
% CONCLUSIONS
%-----------------------------------------------------------------------

\section{Conclusions}

In this paper we have presented the first public release of the open-source and modular weak-lensing measurement, analysis, and
validation pipeline \textsc{ShapePipe}. We have described the objectives and design of the software. In particular, we have emphasised the flexibility of the code architecture, which was designed with the intention of facilitating the inclusion of new processing steps or the improvement of existing ones. This was done with the aim of keeping up with advances made in the field and the possibility of applying \textsc{ShapePipe} to diverse datasets. We have also mentioned steps taken to ensure the reproducibility of results obtained with \textsc{ShapePipe}.

We have provided a brief summary of the tools currently available for masking, source selection, PSF modelling, PSF validation, shape measurement, and calibration. We include links to more detailed documentation for each of these pipeline modules along with examples of how to use them in practice.

Finally, we have mentioned the applications of \textsc{ShapePipe} to real UNIONS data that have been made so far along with plans for future works. We plan to submit follow-up papers in the coming years to inform the community of the advances that have been made in the development of \textsc{ShapePipe} and its applications.

%-----------------------------------------------------------------------
% ACKNOWLEDGEMENTS
%-----------------------------------------------------------------------

\begin{acknowledgements}

This work is based on data obtained as part of the Canada-France Imaging Survey, a CFHT large program of the National Research Council of Canada and the French Centre National de la Recherche Scientifique. Based on observations obtained with MegaPrime/MegaCam, a joint project of CFHT and CEA Saclay, at the Canada-France-Hawaii Telescope (CFHT) which is operated by the National Research Council (NRC) of Canada, the Institut National des Science de l’Univers (INSU) of the Centre National de la Recherche Scientifique (CNRS) of France, and the University of Hawaii. This research used the facilities of the Canadian Astronomy Data Centre operated by the National Research Council of Canada with the support of the Canadian Space Agency. This research is based in part on data collected at Subaru Telescope, which is operated by the National Astronomical Observatory of Japan. We are honored and grateful for the opportunity of observing the Universe from Maunakea, which has the cultural, historical and natural significance in Hawaii. Pan-STARRS is a project of the Institute for Astronomy of the University of Hawaii, and is supported by the NASA SSO Near Earth Observation Program under grants 80NSSC18K0971, NNX14AM74G, NNX12AR65G, NNX13AQ47G, NNX08AR22G, YORPD20\_2-0014 and by the State of Hawaii. This work has made use of the CANDIDE Cluster at the Institut d'Astrophysique de Paris and made possible by grants from the PNCG, CNES, and the DIM-ACAV. This work was supported in part by the Canadian Advanced Network for Astronomical Research (CANFAR) and Compute Canada facilities. We gratefully acknowledge support from the CNRS/IN2P3 Computing Center (Lyon - France) for providing computing and data-processing resources. The authors wish to thank J\'{e}r\^{o}me Bobin, Jean-Charles Cuillandre, Sebastien Fabbro, Joel Gehin, Stephen Gwyn, Fran\c{c}ois Lanusse, Alan McConnachie, Fred Maurice Ngol\'{e} Mboula, Erin Sheldon, Isaac Spitzer and Florent Sureau for various contributions made during the development of this package.

\end{acknowledgements}

%-----------------------------------------------------------------------
% REFERENCES
%-----------------------------------------------------------------------

\bibliographystyle{aa}
\bibliography{ref}

%-----------------------------------------------------------------------
% APPENDIX
%-----------------------------------------------------------------------

\begin{appendix}

\section{Third-party software}
\label{sec:software}

The principal third-party software packages used in \textsc{ShapePipe} are shown in Table~\ref{tab:software}. It should be noted that we are currently using a GitHub fork of \textsc{NGMIX}\footnote{\url{https://github.com/aguinot/ngmix}} for technical reasons.

Where possible, we provide references to the papers in which the software is described in detail. More information concerning the \textsc{cdsclient} package is available at the Strasbourg astronomical Data Center website\footnote{\url{http://cdsarc.u-strasbg.fr/doc/cdsclient.html}}, and information on the \textsc{sqlitedict} package is available at the GitHub repository\footnote{\url{https://github.com/RaRe-Technologies/sqlitedict}}.

\begin{table}[H]
\caption{List of third-party software packages used in ShapePipe.}
\label{tab:software}
\centering
\resizebox{\columnwidth}{!}{
\begin{tabular}{l l p{0.4\linewidth}}
 \hline\hline
 Package name & Fixed version & References\\
 \hline
 \textsc{Astropy} & 5.0 & \citet{astropy:2013,astropy:2018}\\
 \textsc{cdsclient} & 3.84 & \\
 \textsc{GalSim} & 2.2.5 & \citet{rowe:15}\\
 \textsc{Joblib} & 1.1.0 & \citet{joblib:20}\\
 \textsc{Matplotlib} & 3.5.1 & \citet{Hunter:07}\\
 \textsc{MCCD} & 1.2.1 & \citet{liaudat:21}\\
 \textsc{ModOpt} & 1.6.0 & \citet{farrens:20}\\
 \textsc{mpi4py} & 3.1.3 & \citet{dalcin:05,dalcin:08,dalcin:11}\\
 \textsc{NGMIX} & 1.3.6 & \citet{sheldon:15}\\
 \textsc{Numpy} & $\geq$1.20 & \citet{harris:20}\\
 \textsc{Pandas} & 1.4.1 & \citet{pandas:10,pandas:20}\\
 \textsc{PSFEx} & 3.21.1 & \citet{bertin:11}\\
 \textsc{SExtractor} & 2.25.0 & \citet{bertin:96}\\
 \textsc{sip$\_$tpv} & 1.1 & \citet{shupe:12}\\
 \textsc{sqlitedict} & 2.0.0 & \\
 \textsc{TreeCorr} & 4.2.6 & \citet{jarvis:04}\\
 \textsc{WeightWatcher} & 1.12 & \citet{marmo:08}\\
 \hline
\end{tabular}}
\end{table}

\newpage
\section{ShapePipe modules}
\label{sec:modules}

All of the modules currently available in \textsc{ShapePipe} that make up the weak-lensing pipeline described in Sect.~\ref{sec:wl_pipeline} are shown in Table~\ref{tab:modules}. In-depth details on each module are available in the \textsc{ShapePipe} documentation\footnote{\url{https://CosmoStat.github.io/shapepipe/}}.

\begin{table}[H]
\caption{List of ShapePipe modules.}
\label{tab:modules}
\centering
\resizebox{\columnwidth}{!}{
\begin{tabular}{l l p{0.4\linewidth}}
 \hline\hline
 Module name & Fixed Version & Brief description\\
 \hline
 \texttt{find$\_$exposures} & 1.1 & Identify CFIS exposures used to build co-add tiles.\\
 \texttt{get$\_$images} & 1.1 & Download CFIS images from VOSpace.\\
 \texttt{make$\_$cat} & 1.1 & Prepare the final shear catalogue.\\
 \texttt{mask} & 1.0 & Mask out bright stars and other artifacts.\\
 \texttt{match$\_$external} & 1.1 & Match an external catalogue to objects identified with \textsc{SExtractor}.\\
 \texttt{mccd$\_$fit} & 1.1 & Run \textsc{MCCD} to fit a PSF model.\\
 \texttt{mccd$\_$fit$\_$val} & 1.1 & Run and validate the \textsc{MCCD} PSF model.\\
 \texttt{mccd$\_$interp} & 1.0 & Interpolate the \textsc{MCCD} PSF model to galaxy positions.\\
 \texttt{mccd$\_$plotsv} & 1.1 & Produce plots for analysing the \textsc{MCCD} PSF model.\\
 \texttt{mccd$\_$preprocessing} & 1.1 & Pre-process inputs before running \textsc{MCCD}.\\
 \texttt{mccd$\_$val} & 1.1 & Validate the \textsc{MCCD} PSF model.\\
 \texttt{merge$\_$headers} & 1.1 & Merge the headers of split CCD files.\\
 \texttt{merge$\_$sep$\_$cats} & 1.1 & Merge separate catalogues.\\
 \texttt{merge$\_$starcat} & 1.1 & Merge PSF model star catalogues.\\
 \texttt{ngmix} & 0.0.1 & Run \textsc{NGMIX} to measure galaxy shapes and perform metacalibration.\\
\texttt{pastecat} & 1.1 & Combine multiple \textsc{SExtractor} output catalogues into a single file.\\
 \texttt{psfex$\_$interp} & 1.1 & Interpolate the \textsc{PSFEx} PSF model to galaxy positions. \\
 \texttt{psfex} & 1.0 & Run \textsc{PSFEx} to build a PSF model.\\
 \texttt{random$\_$cat} & 1.1 & Create a random catalogue.\\
 \texttt{setools} & 1.1 & Compute summary statistics, produce validation plots, and other useful tools. \\
 \texttt{sextractor} & 1.0.1 & Run \textsc{SExtractor} to extract stars and galaxies.\\
 \texttt{split$\_$exp} & 1.1 & Split single-exposure images into separate files by CCD.\\
 \texttt{spread$\_$model} & 1.1 & Run the spread model method to refine the galaxy sample.\\
 \texttt{uncompress$\_$fits} & 1.1 & Uncompress FITS image files and save them as a single-HDU FITS file.\\
 \texttt{vignet$\_$maker} & 1.1 & Create postage stamps around galaxy positions.\\
 \hline
\end{tabular}}
\end{table}

\end{appendix}

%-----------------------------------------------------------------------
% END OF DOCUMENT
%-----------------------------------------------------------------------

\end{document}